\input{aipcheck}

\documentclass[
    ,final            
  ]
  {aipproc}

\layoutstyle{8x11single}

\begin{document}

\title{Advanced Reconstruction Strategies for the Auger Engineering Radio Array}


\author{Qader Dorosti Hasankiadeh$^\ast$ for the Pierre Auger
Collaboration$^\dagger$}{
  address={$^\ast$Institut f\"ur Kernphysik, Karlsruhe Institute of Technology (KIT), Germany \\
     $^\dagger$Pierre Auger Observatory, Av. San Mart\'in Norte 304,
5613 Malarg\"ue, Argentina \\ (Full author list:
\url{http://www.auger.org/archive/authors_2014_09.html})}
}

\begin{abstract}
The Auger Engineering Radio Array (AERA) aims to detect extensive air showers caused by the interactions of ultra-high energy cosmic rays with the Earth's atmosphere, providing complementary information to the Auger surface, fluorescence and muon detectors. AERA, currently consisting of 124 radio stations, comprises an area of about 6 km$^{2}$. The main objective for exploiting a radio detector is to measure the fundamental air-shower parameters, such as the direction, energy and composition. We have developed reconstruction strategies and algorithms to precisely measure the air-shower parameters with high efficiency. In addition, we will present the results obtained by applying the reconstruction strategies on the experimental data taken by AERA.
\end{abstract}

\maketitle


\section{Introduction}

High-energy cosmic rays impinging onto the atmosphere of the Earth induce cascades of secondary particles, so-called extensive air showers. High fractions of such particles consist of electrons and positrons. As the relativistic particles move in the extensive air shower, two possible radio emission mechanisms can take place: a) the time-variation of the transversal separation of the electrons and positrons due to the geomagnetic field can cause the emission of radiation, i.e \textit{geomagnetic effect}. b) the variation of the negative charge excess along the longitudinal development of the extensive air shower causes radiation, i.e \textit{Askaryan effect}. Such radiation is emitted in a large frequency bandwidth extending over tens of MHz. Being sensitive to the longitudinal air shower, radio detectors can furnish a precise measurement of the mass composition of the primary cosmic rays. In addition, radio detectors provide a quasi-calorimetric measurement of the shower energy, with a duty cycle of about 100\%.    

AERA~\cite{AERA} is the radio extension of the Pierre Auger Observatory~\cite{Auger}, located in the province of Mendoza, Argentina. Using the proven techniques from the previous radio detectors like LOPES~\cite{LOPES} and CODALEMA~\cite{CODALEMA}, AERA aims to measure the fundamental parameters of extensive air showers, such as the direction, energy and mass of the primary cosmic rays that initiated each shower, and covers energies above $10^{17}$ eV. The configuration of AERA-124 consists of 124 radio stations, completed in May 2013, which covers an area of about 6 km$^{2}$. The radio stations consist of two crossed antennas aligned in the geomagnetic north-south and the east-west directions (see fig. 1 in~\cite{butterfly}). AERA records radio signals in the frequency range from 30 to 80 MHz. The detector operates in self trigger and external trigger modes. The external trigger relies on the trigger being sent from Auger's Surface Detector (SD) or Florescence Detector (FD). 

In this proceeding, we present the performance of the reconstruction strategies developed for AERA, using the first set of externally triggered events. For the data analysis, we exploit the radio extension of a partly open source analysis 
framework of the Pierre Auger Collaboration, named Offline~\cite{Offline}. 

\section{Reconstruction of air-shower parameters}
Reconstruction of the fundamental air shower parameters relies on a clear 
distinction of the radio pulses induced by air showers from the pulses caused by 
sources of radio frequency interference (RFI). We present the techniques which
have been exploited to select the signal of radio pulses and reconstruct the shower parameters. 
 
\subsection{Direction reconstruction}
When an event is externally triggered 
by the SD detector, AERA stations are read out. Therefore, there is a 
significant probability that some of the AERA stations contain only 
noise pulses. In order to suppress the contaminations of the noise pulses in the 
AERA stations, we search for a radio signal in an expected time window, exploiting the 
geometrical information of the SD reconstructed shower. The spread of the signal search window
is tuned to be the quadratic sum of the uncertainty of SD direction reconstruction and
the AERA timing precision. Even though the imposed signal search window suppresses 
noise pulses by a large factor, some noise pulses, i.e. false positive pulses, 
fall accidentally into the signal search window. The inclusion of radio stations 
with a false positive pulse can significantly deteriorate the precision of direction
reconstruction, if such (so called isolated) stations stay apart from the cluster of stations with true positive pulses. 
In order to suppress the false positive pulses, we have developed a cluster algorithm to select the cluster of stations 
from the isolated stations. 

\begin{figure}
\centering
\includegraphics[width=0.45\textwidth]{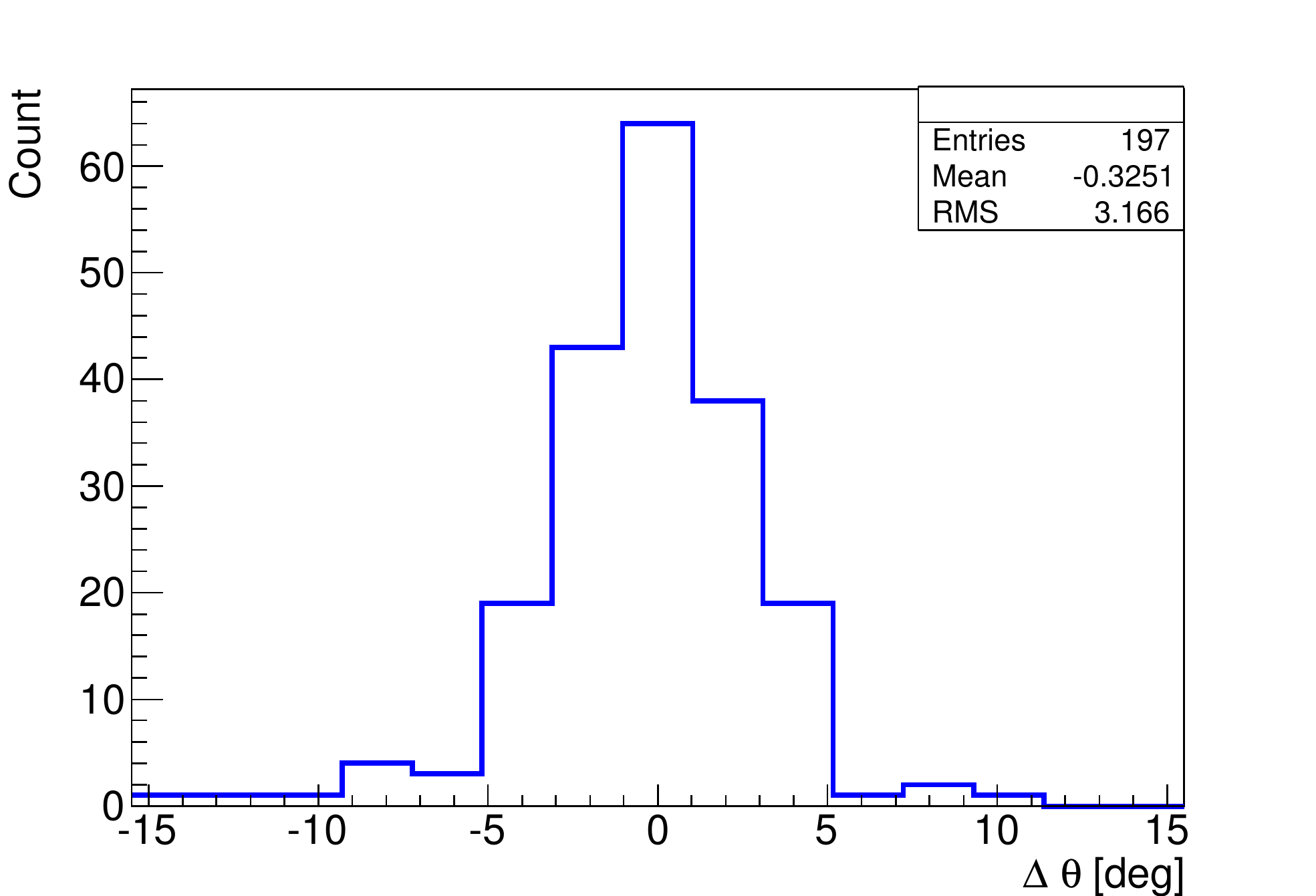}
\includegraphics[width=0.45\textwidth]{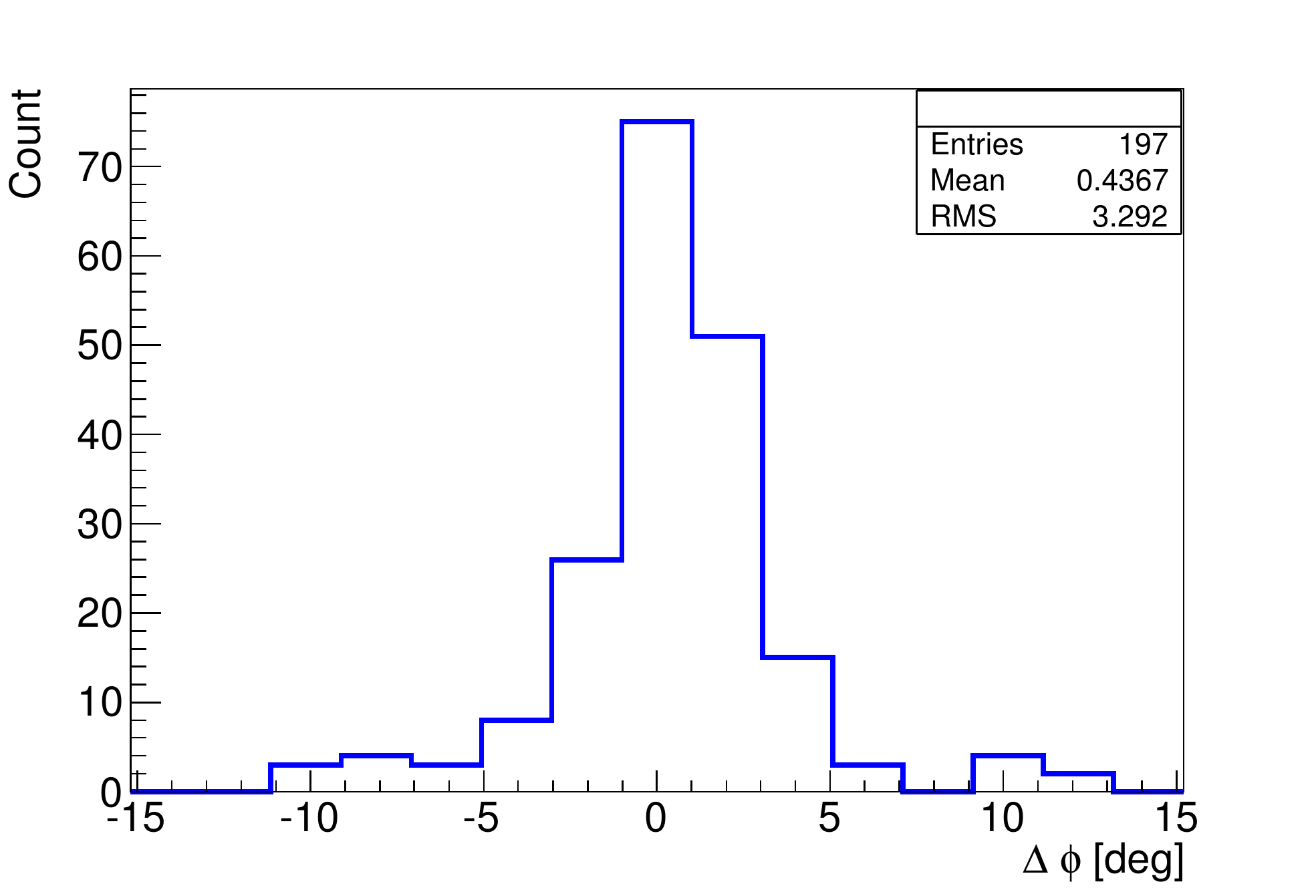}
\caption{The performance of the radio reconstruction is shown through the differences between the radio-reconstructed  and the SD-reconstructed zenith angles  $\Delta \theta$ (left) and azimuth angles $\Delta \phi$ (right) for each shower.}
\label{fig:Direction}
\end{figure}

The performance of the radio direction reconstruction is shown in Fig.~\ref{fig:Direction}. 
The radio direction reconstruction agrees with the SD direction reconstruction 
better than 4$^{\circ}$. The dataset used in this analysis contains mostly low-energy events which 
triggered only few stations per events. Therefore, there is much room left for improvements when analysing higher-energy events.

\subsection{Energy reconstruction}

\begin{figure}
\centering
\includegraphics[width=0.48\textwidth]{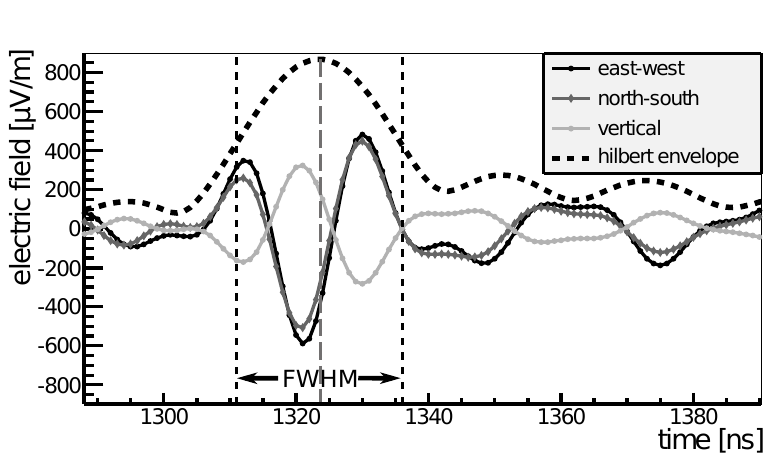}
\includegraphics[width=0.35\textwidth]{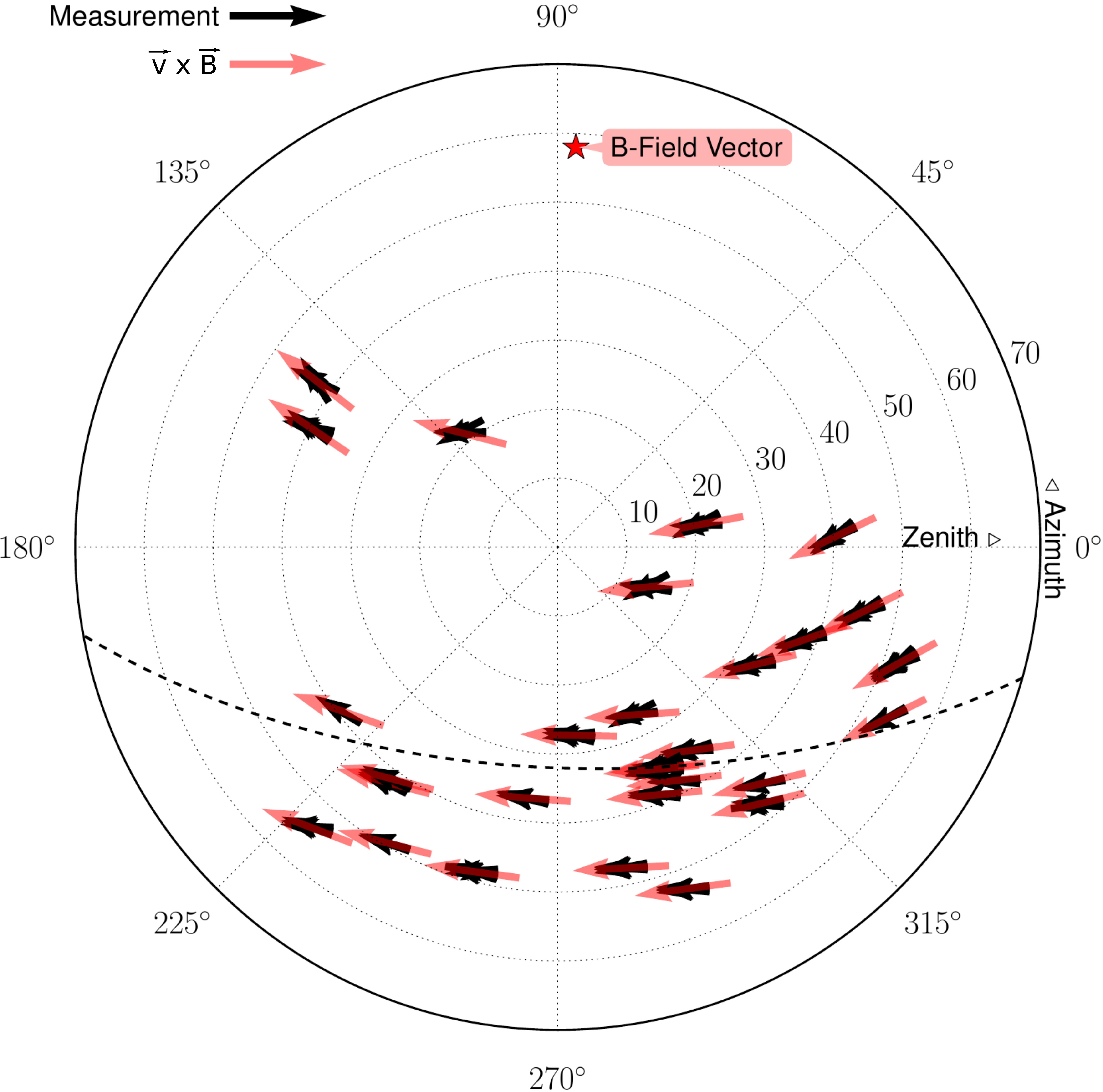}
\caption{Left) time trace of a reconstructed electric field vector for east-west (solid black line), north-south (solid dark grey line)
and vertical (solid light grey line) directions. The Hilbert transformation is
applied to calculate the envelopes of the time trace of the electric field vector. The dashed line shows the 
Hilbert envelope. The figure is adapted from~\cite{Glaser}. Right) arrival direction of externally triggered events for which the measured electric field (dark black arrows) is compared with the expected electric field induced by geomagnetic mechanism (light red arrows) . The figure is adapted from~\cite{Klaus}.}
\label{fig:energyEstimator}
\end{figure}

Our energy reconstruction algorithm follows the idea that the strength 
of the electric field vector of an air shower at the optimal distance $D_{0}$, corrected
for the angular dependency of the geomagnetic emission mechanism, is correlated to the 
energy of the air shower~\cite{Glaser}. The electric field vector can be reconstructed from 
the voltage measurements registered by each radio station. Each radio station measures 
voltages in two polarisation directions, i.e. north-south and east-west. Therefore, by exploiting 
the reconstructed direction of an air shower and correcting for the electronic responses of the antennas, 
we can reconstruct the electric field vector. 
The Hilbert transformation is used to calculate the envelopes of the time traces of each
component of the electric field vector (see fig.~\ref{fig:energyEstimator}).
Then the strength of the electric field vector is defined as the square root of the maximum of the 
quadratic sum of the Hilbert envelopes. 

The geomagnetic emission is the dominant emission mechanism at the AERA site~\cite{polarisation}. 
The electric field induced by geomagnetic effect points to the direction of $\vec{v}\times \vec{B}$, 
where $\vec{v}$ is a vector in the direction of the shower and $\vec{B}$ is the Earth magnetic field.
The strength of such an emission mechanism depends on the angle $\alpha$ between 
the shower axis and the Earth magnetic field. We thus correct for the geomagnetic 
emission mechanism, dividing the measured strength of the electric field $\left | \vec{E} \right |$ by $\sin \alpha$.

Fig.~\ref{fig:energyEstimator} (right) compares the directions of measured electric field vectors with the expected electric field induced by geomagnetic field, for some externally triggered events. There is a good agreement between the measured and expected electric fields. Thus we define an energy estimator exploiting the strength of the electric field vector at an optimal distance of $D_{0}$. The value of $D_{0}$, i.e. 110 m, is optimised to achieve the highest energy resolution for AERA~\cite{Glaser}. 

A preliminary test on 100 externally triggered events shows that the energy estimator can describe the SD energy measurements with a deviation less than $30 \%$.

\subsection{Mass composition reconstruction}

Radio detectors have provided proven techniques to reconstruct the composition of cosmic rays. 
For instance, it has been observed that the slope of the lateral distribution of
measured radio signals with LOPES is correlated to the mass composition of cosmic rays~\cite{IDK, NunziaLDF}. 
In addition, we have evidences that the shape of the radio wavefront is correlated to the 
mass composition~\cite{FrankWF}. The primary mass of cosmic rays is strongly correlated with the depth of air shower maximum, i.e. the atmospheric depth of an air shower in which the number of secondary particles is maximised. It is believed that
most of the contribution of radio emissions is from the region around and before the shower maximum~\cite{marianne}, therefore,
 a radio observable, e.g. the radio wavefront, carries the information about the geometrical distance $L_{max}$ from ground to shower maximum. 
In this section, we investigate the relation between the shape of the
radio wavefront and $L_{max}$, using some AERA simulations and experimental data.

Recent observations from LOFAR reveal that the radio wavefront has a hyperbolic shape~\cite{FrankWF, LOFAR}.
However, a hyperbola can be approximated as the cone asymptotes approaching the hyperbola at large distances~\cite{FrankWF}.
The opening angle $\rho$ of the cone is correlated to $L_{max}$.  

In this study, we used a set of Monte-Carlo (MC) data in order to parametrise the $L_{max}$ parameter.
The MC dataset consists of 720 proton primaries with energies ranging from 0.28 EeV to 1 EeV and 
zenith angles ranging from 30$^{\circ}$ to 50$^{\circ}$.

\begin{figure}
\centering
\includegraphics[width=0.45\textwidth]{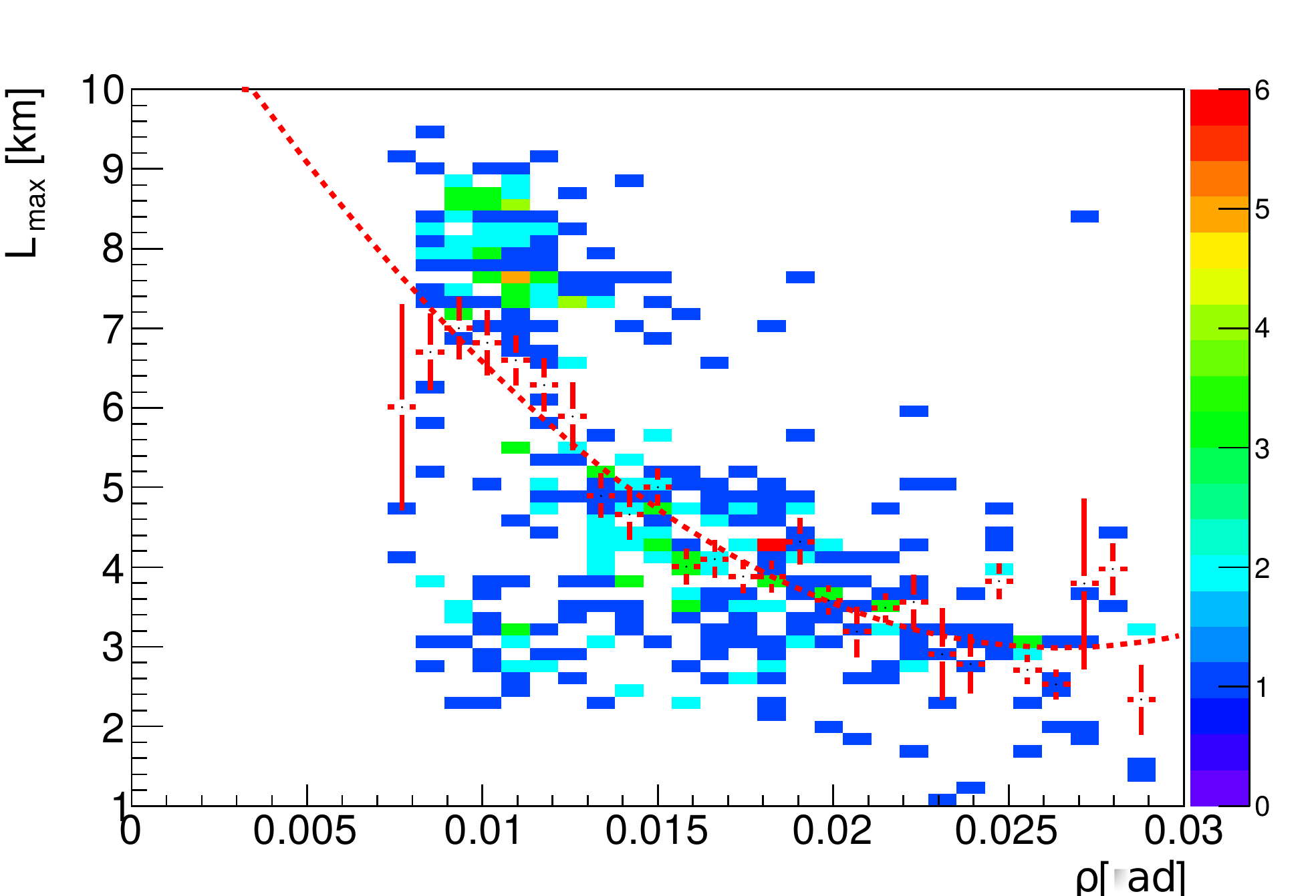}
\includegraphics[width=0.45\textwidth]{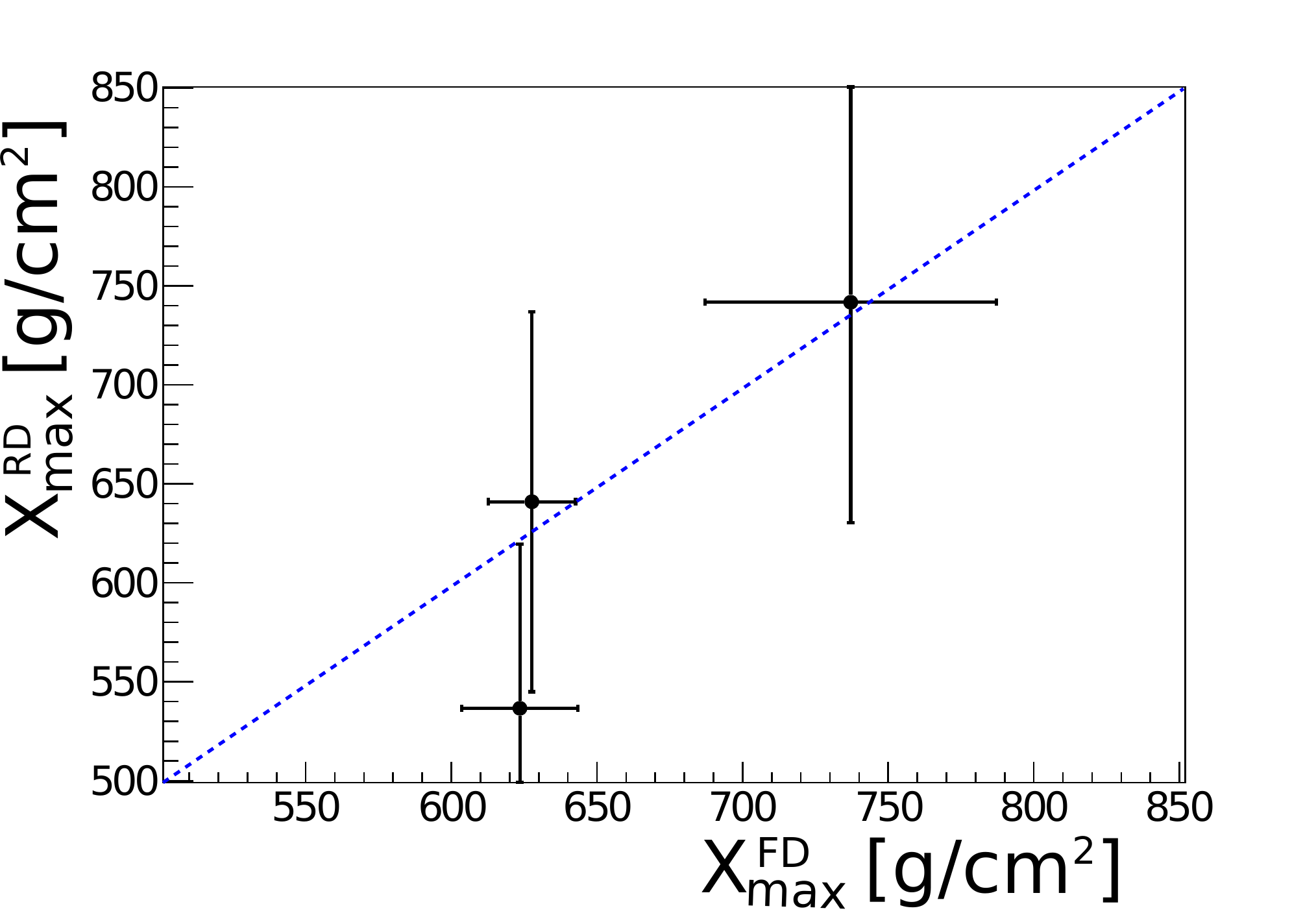}
\caption{Left) shows the distribution of distance $L_{max}$ to shower maximum as a 
function of the cone opening angle $\rho$ for the MC dataset. The dashed line shows
the polynomial function fitted on the profile distribution. Right) presents the depth of shower maximum reconstructed with AERA, $X_{max}^{RD}$,
 as a function of that reconstructed with Auger florescence detector, $X_{max}^{FD}$.
The dashed line illustrates the 1:1 ratio of the reconstructed parameters. }
\label{fig:correlation}
\end{figure}

Fig.~\ref{fig:correlation} (left) shows the distribution of the distance $L_{max}$ to shower maximum as a 
function of the cone opening angle $\rho$ for the MC dataset. In order to parametrise $L_{max}$, a polynomial is fitted 
on the profile of the distribution. Fig.~\ref{fig:correlation} (right) presents the distribution of the depth of shower maximum reconstructed with AERA, $X_{max}^{RD}$,
 as a function of that reconstructed with Auger fluorescence detector, $X_{max}^{FD}$, for three coincident events. The error bars correspond 
to the fit uncertainty used for the parametrisation of $L_{max}$ and the measurement uncertainty of the 
FD. There is a hint for a correlation between the reconstructed shower maximum of AERA and FD.  

\section{Conclusion}

AERA is the radio extension of the Pierre Auger Observatory. The AERA-124 configuration was
completed in 2013, providing an instrumentation area of about 6 km$^{2}$. Reconstruction of the fundamental air shower parameters,
such as direction, energy and primary mass, relies on the identification
 of the radio pulses caused by air showers from the pulses caused by RFI. We have 
developed strategies to select the radio pulses and reconstruct the shower parameters. 
We perform a comparison between the air shower parameters measured by AERA and 
those measured by the surface and florescence detectors of the Pierre Auger Observatory. The results are:
\begin{itemize}
\item  The direction of air showers measured by AERA agrees better than 4$^{\circ}$ with that measured by SD, leaving much
rooms for improvement for high-energy events for which the number of triggered stations are higher.
\item The relative difference between the SD-reconstructed energy and radio-reconstructed energy is 
better than 30$\%$.
\item There is a hint of a correlation between the 
shower maximum reconstructed from the radio wavefront measured by AERA and 
the one measured by FD.
\end{itemize}



\bibliographystyle{aipproc}   

\bibliography{sample}

\IfFileExists{\jobname.bbl}{}
 {\typeout{}
  \typeout{******************************************}
  \typeout{** Please run "bibtex \jobname" to optain}
  \typeout{** the bibliography and then re-run LaTeX}
  \typeout{** twice to fix the references!}
  \typeout{******************************************}
  \typeout{}
 }

\end{document}